\documentclass[titlepage]{article}
\usepackage[usenames,dvipsnames,svgnames,x11names]{xcolor}

\newcommand{\para}[1]{\noindent \textbf{#1.}}

\usepackage[normalem]{ulem} 

\usepackage{bbold} 
\usepackage[sans]{dsfont}

\usepackage[nocompress]{cite}
\usepackage{listings}

\lstdefinelanguage{XML}
{
basicstyle=\ttfamily\footnotesize,
  morestring=[b]",
  moredelim=[s][\bfseries\color{Maroon}]{<}{\ },
  moredelim=[s][\bfseries\color{Maroon}]{</}{>},
  moredelim=[l][\bfseries\color{Maroon}]{/>},
  moredelim=[l][\bfseries\color{Maroon}]{>},
  morecomment=[s]{<?}{?>},
  morecomment=[s]{<!--}{-->},
  commentstyle=\color{gray},
  stringstyle=\color{blue},
  identifierstyle=\color{red}
}
\usepackage{moreverb}

\usepackage[nounderscore]{syntax}

\usepackage[pdftex]{graphicx}
\graphicspath{{./figures/}}
\DeclareGraphicsExtensions{.pdf}
\usepackage[cmex10]{amsmath}
\usepackage{amssymb}
\usepackage{mathtools}
\usepackage{amsfonts}

\usepackage{subfig}
\usepackage{algorithmicx}
\usepackage{algpseudocode}
\usepackage[ruled]{algorithm}
\definecolor{light-gray}{gray}{0.75}
\algrenewcommand{\algorithmiccomment}[1]{\hskip3em{{\footnotesize \textcolor{light-gray}{$\blacktriangleright$}}} #1}
\usepackage{url}
\usepackage{multirow} 
\usepackage{rotating} 
\usepackage{booktabs} 
\usepackage{colortbl} 
\usepackage{tablefootnote} 

\usepackage[pdftex,colorlinks=true,urlcolor=blue,citecolor=blue]{hyperref}

\usepackage{xspace}

\usepackage{enumitem}

\newcommand{\echo}{$\mathbb{ECHO}$\xspace}
\newcommand{\echob}{$\mathds{ECHO}$\xspace}

\usepackage{blindtext}

\begin{document}

\title{\echob: An Adaptive \underline{O}rchestration Platform for \underline{H}ybrid Dataflows across \underline{C}loud and \underline{E}dge}
%
%
\author{Pushkara Ravindra, Aakash Khochare, Siva Prakash Reddy,\\Sarthak Sharma, Prateeksha Varshney and Yogesh Simmhan\\
~\\
\emph{EMail: pushkar1593@gmail.com, simmhan@cds.iisc.ac.in}\\
\url{https://github.com/dream-lab/echo}}
%
%
%

\maketitle              


\begin{abstract}
The Internet of Things (IoT) is offering unprecedented observational data that are used for managing Smart City utilities. \emph{Edge} and \emph{Fog} gateway devices are an integral part of IoT deployments to acquire real-time data and enact controls. Recently, \emph{Edge-computing} is emerging as first-class paradigm to complement Cloud-centric analytics. But a key limitation is the lack of a platform-as-a-service for applications spanning Edge and Cloud. Here, we propose \echo, an orchestration platform for dataflows across distributed resources. \echo's hybrid dataflow composition can operate on diverse data models -- streams, micro-batches and files, and interface with native runtime engines like TensorFlow and Storm to execute them. It manages the application's lifecycle, including container-based deployment and a registry for state management. \echo can schedule the dataflow on different Edge, Fog and Cloud resources, and also perform dynamic task migration between resources. We validate the \echo platform for executing video analytics and sensor streams for Smart Traffic and Smart Utility applications on Raspberry Pi, NVidia TX1, ARM64 and Azure Cloud VM resources, and present our results.
\end{abstract}

\section{Introduction}
%
The growth of Internet of Things (IoT) is leading to an unprecedented access to observational data about physical infrastructure such as traffic/surveillance cameras and smart power meters in Smart Cities, as well as social life-style through fitness bands like FitBit and automation assistants like Google Home. Such data streams are integrated with historic data and analytics models to make intelligent decisions, such as managing traffic signaling or power grid optimization in cities~\cite{simmhan:cise:2012,amrutur:lightpole}, or controlling devices in your home.

%
Traditionally, all this decision making and analytics have taken place in the Cloud due to their easy service-oriented access to seemingly infinite resources. Data is streamed from the edge devices and sensors to the data center, and control decisions communicated back from the Cloud analytics to the edge for enactment. This, however, has several down-sides. The \emph{network bandwidth} to send high-fidelity video streams to the Cloud can be punitive, and the round-trip \emph{latency} required to move data from the edge to the Cloud and control signals back can be high. Clouds also have a pay-as-you-go model where data transfers, compute, and storage are all \emph{billed}~\cite{simmhan:iot:2017,epema:edge}. 

An integral part of IoT deployments are \emph{Edge and Fog devices} that serve as gateways to interface with sensors and actuators on the field. These are typically collocated or within few network hops of the sensors, and have non-trivial compute capacity. E.g., a Raspberry Pi 2B device, popular on the Edge, has 4 power-efficient ARM cores, each performing at about $\frac{1}{3}^{rd}$ an Intel Xeon E5 core on the Cloud~\cite{ghosh:2016}. Devices like the NVIDIA TX1 and Softiron ARM64 servers offer accelerators and energy-efficiency that can be ruggedized for deployment as a Fog layer. Rather than just have them move data and control signals between the field devices and the Cloud, these Edge and Fog resources should be actively considered as first-class computing platforms to complement the Cloud-centric model to reduce the network transfer time and costs~\cite{epema:edge,varshney:icfec:2017}. There is also the lost opportunity of not using their \emph{captive computation} capability.

There have been \emph{ad hoc} or custom applications that indeed leverage Edge, Fog and Cloud resources together. However, a key hurdle to adoption of this distributed paradigm is the lack of a platform ecosystem that simplifies the composition, deployment, and management of applications, micro-services and data seamlessly across these computing layers. In this regard, we are in a situation similar to feature phones before smart phones came along, where middleware has not kept up with hardware and communication advances~\cite{simmhan:iot:2017}. In this article, we highlight key requirements for such a distributed orchestration platform to support the novel requirements of IoT applications on diverse resources, reaffirming earlier works~\cite{DBLP:journals/corr/MineraudMST15a}. We further propose \echo, an architecture and platform implementation that addresses these needs.

%
There are existing solutions in the commercial and open source community that partially address this gap. Amazon's Greengrass and Microsoft's Azure IoT Edge provide gateway management SDK that tightly integrates with their Cloud services~\cite{greengrass,azureiot}. Eclipse Kura and Liota~\cite{kura,liota} are gateway management services which support local applications, while platforms like Edgent, Node.RED, and NiFi support basic dataflow capabilities that are limited to stream or micro-batch data. Our work goes beyond these offerings 
and examines hybrid data models (stream, micro-batch, batch), generic dataflow composition, pluggability with external platforms (TensorFlow, Storm, Spark), and dynamic migration.

%
Specifically, we make the following contributions in this paper:
\begin{itemize}
\item We highlight the key features and desiderata for a platform to support distributed application composition and execution across Edge, Fog and Cloud devices (\S~\ref{sec:requirements}).
\item We propose \echo, an architecture and open source platform for computing across Edge and Cloud that meets these requirements, while also leveraging existing open source tools (\S~\ref{sec:arch}).
\item We validate \echo for several representative Smart City applications, including video, stream and event analytics (\S~\ref{sec:results}).
\end{itemize}

Besides these contributions, we also review related literature in \S~\ref{sec:related}, and present our conclusions and future work in \S~\ref{sec:conclusions}.

\section{Requirements and Motivation}
\label{sec:requirements}

\begin{figure*}[t]
   \centering
   \includegraphics[width=0.95\textwidth]{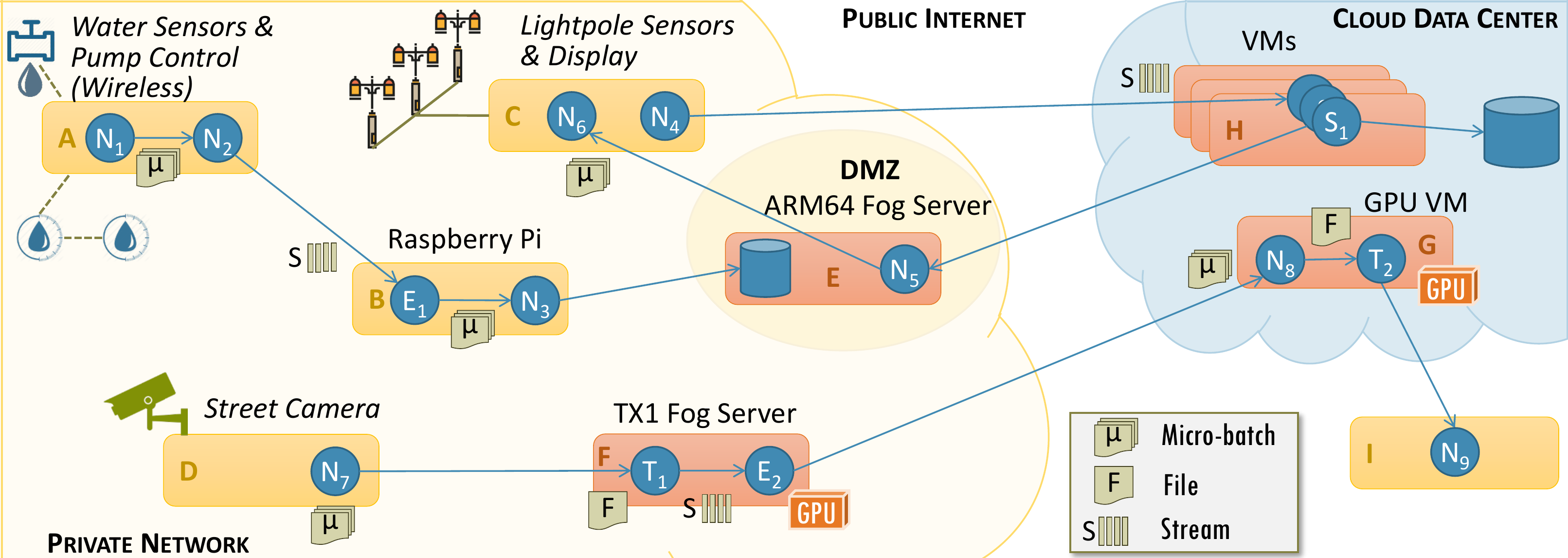}
   \caption{Motivating Usecase from a Smart Community in a City}
   \label{fig:motive}
\vspace{-0.15in}
\end{figure*}
Fig.~\ref{fig:motive} illustrates scenarios for a \emph{Smart Community}, where sensors and actuators like water level and quality sensors and pump controls for \emph{smart water ma\-nage\-ment}, environment sensors and digital displays fixed on street light poles for ambient \emph{urban sensing and public notification}, and PTZ cameras for \emph{surveillance and traffic} are present~\cite{amrutur:lightpole}. Edge devices like Raspberry Pi and smart phones, NVidia TX1 and ARM64 Fog servers, along with Cloud VMs, are present in the private (community and Cloud) networks, and the public Internet, for executing analytics and storage. This motivates several key and distinct requirements for an IoT platform that allows composition and execution of decision making applications across Edge, Fog and Cloud resources, as we discuss below.

\para{Dataflow Composition Model} Data-driven IoT applications are well-suited for a dataflow programming model, where user tasks are vertices in a \emph{directed acyclic graph (DAG)} that execute upon data arrival, 
and edges are channels that route the data between tasks. Many Big Data platforms like \emph{Apache Spark, Storm} and \emph{Google's TensorFlow}, and edge-centric platforms like \emph{Edgent} and \emph{MiNiFi} use a dataflow model. It also allows a library of tasks to be developed and reused by diverse domains, and these tasks form the unit of scheduling on compute resources. E.g., Fig.~\ref{fig:motive} shows tasks $N_1,N_2,E_1$ and $N_3$ tasks operating as a linear dataflow on water events that are processed and stored to a database.

\para{Hybrid Data Sources} IoT applications often operate over thousands of observation streams, performing low-latency event pattern detection, e.g., on water event streams at $E_1$ in Fig.~\ref{fig:motive}. We also require batch processing on accumulated data for high throughput, say for traffic mining over video segments by $T_1$. Micro-batches, like from $N_1$ to $N_2$, offer a stream of batched tuples, balancing latency and throughput. Hence, \emph{seamlessly allowing hybrid datasets} to pass between tasks in the dataflow is essential, allowing the application composer to select the appropriate data model. 
Lambda Architecture and platforms like Flink and Spark Streaming affirm the need for such hybrid models~\cite{kiran2015lambda}. 
This also affects the 
\emph{QoS} for the dataflows (e.g., latency, throughput, reliability, price).

\para{Diverse Resource Capabilities} Edge, Fog and Cloud resources have heterogeneous capabilities. 
Platforms like Pi and Arduino are popular as edge devices (e.g., a Pi 2B with 1~GHz CPU/1~GB RAM running Linux, costing US\$~35). IoT Fog servers from vendors like Dell and NVIDIA offer energy efficient multi-core ARM64 processors and GPGPUs 
(e.g., NVidia TX1 with a GPU, Softiron ARM64 server). On-demand Cloud VMs at different globally spread-out data centers are also accessible. The software platform must be able to \emph{leverage such Edge, Fog and elastic Cloud VMs} to meet the application QoS,
while also being aware of constraints like energy (e.g., if powered by battery or solar) and pricing. 

\para{Network Connectivity} 
IoT compute resources are distributed. So the network connectivity between them is crucial. 
The resources may be within \emph{local networks} (e.g., Cloud data center, 
private campus) and \emph{wide-area networks} (e.g., devices across a city), with variability in \emph{bandwidth and latency} ranging from 10-1000~ms and Kbps-Gbps
, 
depending on the medium (3G/WiFi/LoRa). 
Communication within a \emph{private network}, a \emph{public network}, or between the two with firewalls 
also impacts the 
visibility and accessibility of service endpoints. The platform should \emph{transparently resolve this} (e.g., push vs. pull) during dataflow orchestration. 

\para{Native Runtime Engines} Numerous Big Data and emerging edge platforms exist for data processing. Some like Spark and Storm are general purpose, allowing custom logic, while others like Edgent and TensorFlow are specialized for event analytics and deep learning, which are popular in IoT. Packages like R may also require command-line execution. These are also optimized for different resources (e.g., VMs, edge, GPU). The execution platform should \emph{leverage the strengths of native runtime engines} while coordinating between them like a ``meta-engine'' (e.g., data model mapping, public/private networks, scheduling), and also offering basic dataflow orchestration. E.g., in Fig.~\ref{fig:motive} shows the use of \emph{Edgent} ($E_1, E_2$) for complex event processing (CEP) on Pis, \emph{TensorFlow} for classifying image batches using deep neural networks on GPUs ($T_1, T_2$), \emph{Storm} for scalable streaming analysis on Cloud VMs over ambient observations ($S_2$), with \emph{NiFi} as the baseline dataflow orchestrator ($N_1-N_9$). 

\para{Service-Oriented Architecture} Cloud owes its success to its Service-Oriented Architecture (SOA), at the infrastructure (IaaS), platform (PaaS) and software (SaaS) levels.
Edge and Fog platforms can similarly benefit. Infrastructure services at these resource layers can use \emph{containers} like LXC and Dockers for resource  sand-boxing. They are more light-weight than hypervisors and offer fast startup, 
but trade-off strict security with multi-tenancy. 
\noindent \emph{Platform micro-services} 
are viable on constrained edge and Fog layers for rapid dataflow deployment~\cite{pautasso2017microservices}. A platform service on the edge or Fog resource can perform local task coordination and data transfers across resources, and manage  
the application lifecycle.


\para{Discovery and Adaptivity} Decentralized IoT resources operate in a dynamic environment where the availability and capacity of edge and Fog resources can vary over time (e.g., network link, mobility, battery level). 
This is unlike public Clouds that have on-demand and reliable availability. This requires a \emph{scalable registry service} to publish the health metrics of edge and Fog devices, and to track their applications. 
Maintaining the available data 
sources, and dataflows is useful 
when making scheduling decisions, and for provenance and billing. Lastly, the inherent dynamism of the resources, data sources, and applications along with the need to meet QoS for dataflows makes it necessary to support \emph{dynamic migration of dataflows} between different resources as a first class capability.

\section{The \echo Architecture}
\label{sec:arch}

Here, we propose \echo, an adaptive \underline{o}rchestration platform for \underline{h}ybrid dataflows across \underline{C}loud, Fog and \underline{E}dge resources\footnote{\echo is available for download at \url{https://github.com/dream-lab/echo}}. \echo's design addresses the requirements we identify. Next, we discuss the infrastructure and platform abstractions that \echo supports, and then delve into its architecture design (Fig.~\ref{fig:arch}).

	\subsection{Resource Infrastructure}
\echo is designed for resources with diverse capabilities, with a baseline being a Linux device with $\approx$ 1~GHz CPU/1~GB RAM, and able to run \texttt{cgroups} containers and a Java Runtime. Resources themselves may be devices or servers that are \emph{internally managed} by \echo (like edge and Fog devices), or \emph{externally managed} IaaS resources, like on-demand VMs from (public/private) Cloud service providers. 
We have a \emph{Device Service} that acts as an infrastructure fabric to bootstrap and control internally managed resources. It registers the compute, accelerator, memory, disk and network capacity, IP address,  visibility of the device from public or private networks, etc. of the device with a \emph{Registry Service} (discussed later) to make it available. It also periodically reports performance statistics of the device (e.g., CPU\%, Memory\%) for health monitoring.

Internally managed resources use \emph{containers} for application deployment, light-weight resource allocation, sand-boxing applications (and the base device) for limited security, and for billing. We skip this for external resources since the IaaS provider takes these responsibilities. We use \texttt{LXC} containers based on \texttt{cgroups} capability of the Linux kernel, though Docker is also viable but more resource intensive for low-end edge devices. The Device Service starts, stops and manages containers on internal resources, and can deploy the appropriate container image requested for application initiation. The container's lifecycle is also registered with the Registry, along with its periodic performance metrics.  

	\subsection{Programming Model}
\label{sec:model}
	\begin{figure*}[t]
    \centering
   \includegraphics[width=0.95\textwidth]{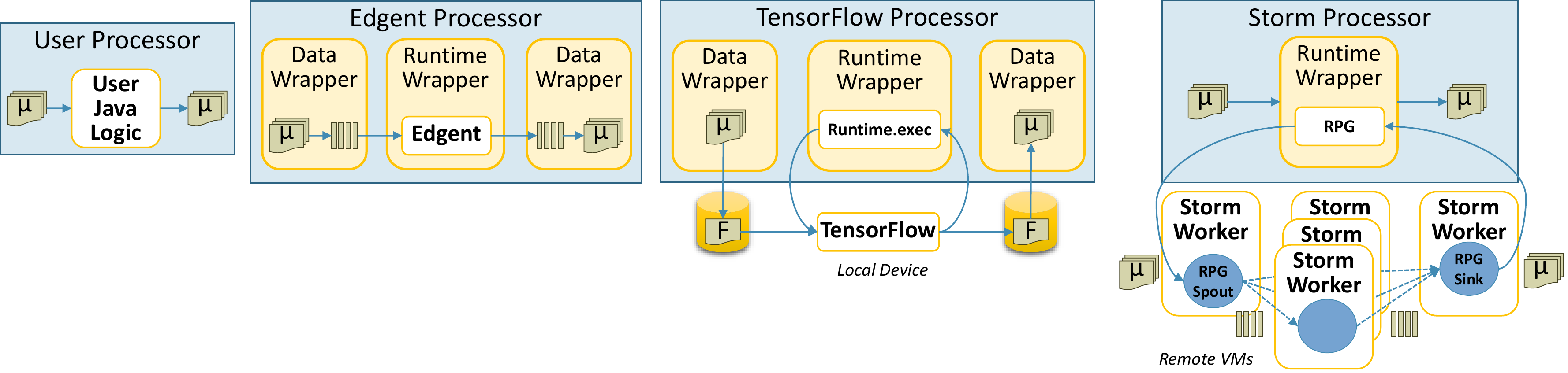}
    \caption{Wrappers in \echo for hybrid data models \& external engines}
    \label{fig:model}
\end{figure*}
\echo adopts a \emph{dataflow programming model} composed as a 
directed graph allowing cycles, which is similar to but more flexible than DAGs
that are widely used in business processes and Big Data applications. Vertices represent tasks (or \emph{processors}) with custom user logic that are executed when an input data item is available, and can generate zero or more output data items. The edges represent the data dependencies and data movement between the tasks. 

Data items consumed and produced by tasks can be of three forms: \emph{streams, files,} or \emph{micro-batches}. Streams have a sequence of unbounded tuples available in-memory, files are a collection of bytes on disk, while micro-batches are a set of tuples or bytes in-memory. User processors are annotated with the data model that they use on their input and output. While we use micro-batch as the default model between processors, \echo automatically maps between the stream or files to/from micro-batch. This is done by \emph{data wrappers} around the task logic that accumulate event streams from tasks into windows to form a micro-batch, and similarly replay events from the micro-batch to the task as a stream (Fig.~\ref{fig:model}). Likewise, micro-batches can be written to and read from the device's file system as files
for passing to the task. This allows users to focus on the business logic and not on data model transformations. 


Lastly, the \echo programming model provides native support for interfacing with external runtime engines using specialized \emph{runtime wrappers} (Fig.~\ref{fig:model}). 
These processors take the native dataflow for an external runtime engine, initialize that engine, pass input data to it, and receive the results back, using data wrappers if needed. Such engines may be in-memory Java libraries, commandline executables, or a remote Big Data platform. Specifically, we support \emph{Apache Edgent}~\cite{edgent}, an in-memory Java CEP engine for edge devices that consumes and produces event streams, and executes online pattern queries on them. A processor for \emph{Google's TensorFlow}~\cite{DBLP:journals/corr/AbadiABBCCCDDDG16} executes classification models as a local Python process, with access to CPU and GPGPU, using file-based input and output. We also support \emph{Apache Storm} and \emph{Spark} platforms on clusters/VMs, using data transfer bindings between a local processor and the remote application.


\begin{figure*}[t!]
    \centering
   \includegraphics[width=0.95\textwidth]{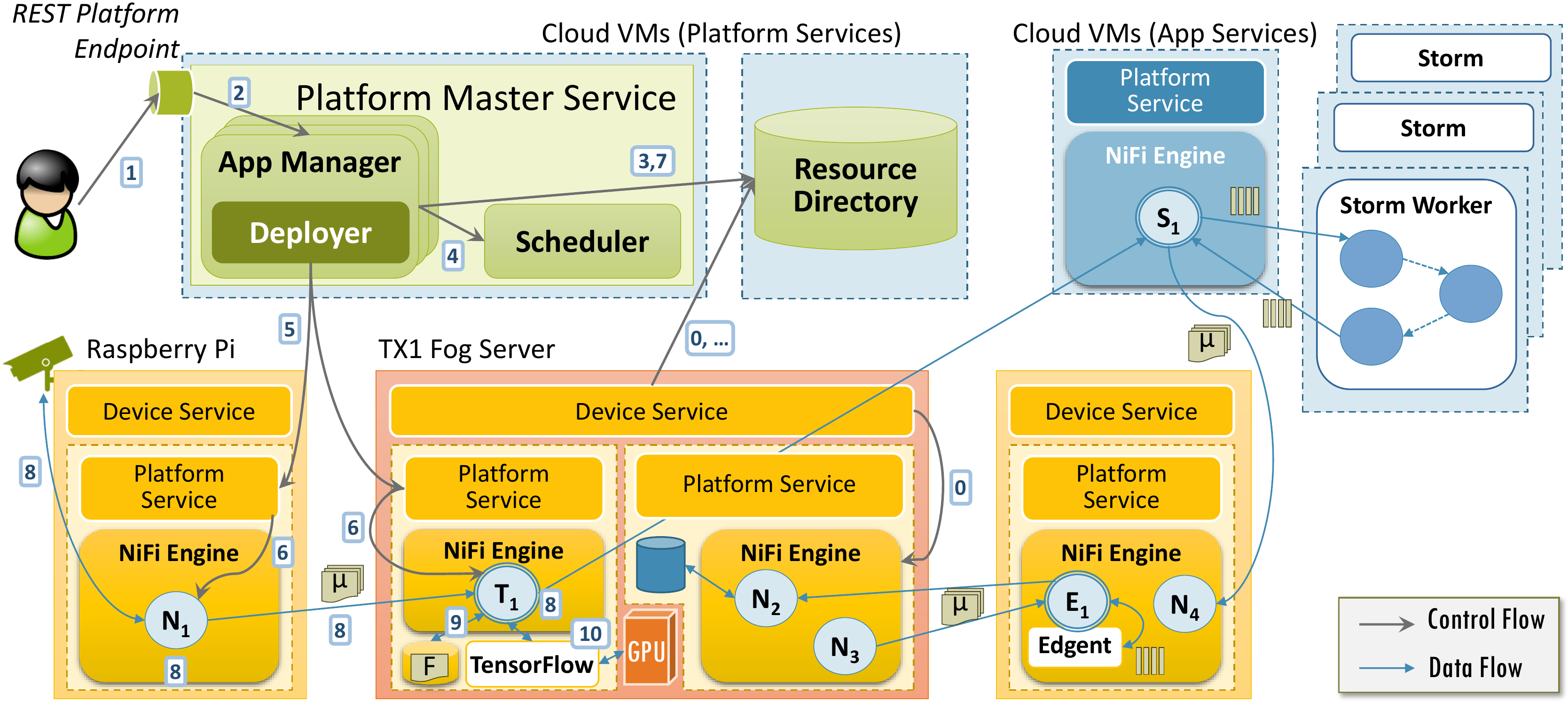}
    \caption{\echo Platform Architecture}
    \label{fig:arch}
\end{figure*}

	\subsection{Platform Design and Implementation}

Fig.~\ref{fig:arch} shows the high level Platform architecture of \echo. Internally managed devices have the \emph{Device Service} running on them as part of the infrastructure fabric. A \emph{Platform Service} runs on each container or VM and interfaces with a local \emph{Apache NiFi} instance which we use as our default dataflow engine. A \emph{Resource Directory} and \emph{Platform Master} form the core platform services, typically hosted on a public Cloud VM. The devices, their containers and externally managed Cloud VMs available for running user dataflows are registered with the Resource Directory. The master is responsible for managing the lifecycle of a dataflow on behalf of the user by coordinating with the other services. Next, we discuss individual components of \echo and their interaction pattern.


\noindent\textbf{Resource Directory.}
The resource directory is a \emph{registry} of all state in the system. We use it to register resources and dataflows but it is naturally suited for data items as well. We use the \emph{Hypercat 3.0} BSI standard~\cite{hypercat3} that has been developed as a light-weight JSON-based registry for IoT and Smart City assets. 
Each registered \texttt{item} is identified using a unique \texttt{href} URI and associated \texttt{item-metadata} which is a list of \texttt{rel}ationship and \texttt{val}ue pairs. Besides relationships like \texttt{description}, 
geolocation, last updated timestamp and event streams, it also allows user-defined relationships. Hypercat exposes REST-based registration (\texttt{POST}) and query (\texttt{GET}) of this JSON including geographical and lexicographic search, subscription to event stream updates, and web-based security. We extend an existing 
Hypercat implementation for our needs~\footnote{\url{https://github.com/HyperCatIoT/node-hypercat}}.

We define a logical hierarchy based on the \texttt{href}'s path with the first level having the type of resource, such as \emph{device} or \emph{dataflow}, the next level having the \emph{unique ID} for the item, and subsequently, sub-categories within that item. E.g., for an edge device, we may have \emph{href=http://tempuri.org/device/e97e0195acf4}, while its CPU usage may be at \emph{href=http://tempuri.org/device/e97e0195acf4/ CPUUtil}. Since the entire JSON entry for an item is updated when even one relationship changes, having such href-based logical grouping allows fine-grained updates and queries.  For devices and containers, we capture information such as the capacity (core, memory, disk, NIC, accelerators), IP address, and the current utilization. 
For dataflows, we capture the JSON of the actual directed graph of processors, their mapping to specific resources, and their state. 
This can be further extended to record the data items generated, sensor events streams available, etc. based on user needs for dynamic binding of dataflows to sources.

The entries in the catalog are populated by the Device Service and the Platform Service when resources come online, with a monitoring thread updating the resource usage. The App Manager inserts and updates the state of the dataflow when it is started, updated, rebalanced or stopped. Besides external services that can use the catalog, the scheduler queries for information on the available resource capacities to match the dataflow processor requirements using prefix and exact search capabilities of Hypercat.

\noindent \textbf{Device Service.} 
The Device Service is an infrastructure service running on internally managed devices that monitors the device and the containers it spawns. It registers the device on bootup, and each container it spins up or shuts down, with the Resource Directory (step $0$ in Fig.~\ref{fig:arch}). The service exposes a REST API that can be used to launch new containers using \texttt{LXC} with specific application images, and turn down unused containers. It also logs the CPU and Memory utilization for each device its containers with the registry. This gives the capacity of the device and also the performance of applications within its containers.

\noindent \textbf{Platform Master and Dataflow Lifecycle.}
The Platform Master is a REST service responsible for managing a dataflow's lifecycle for the user using other \echo components. The master itself is registered with the registry for bootstrap. 
The service exposes three main actions: starting a dataflow, stopping it, and dynamically rebalancing it. These can be easily extended to other variants such as pausing, changing input parameters, or even modifying the structure of the dataflow. Fig.~\ref{fig:arch} 
illustrates a dataflow starting. 
Users \texttt{POST} a composed dataflow JSON to the master service, which spawns an \emph{App Manager} thread to handle the request for this dataflow. The master is designed to be stateless, with all state managed in the registry. The manager queries the registry for the available resources -- registered containers or VMs and their current capacity, which it passes to the \emph{Scheduler} along with the dataflow. The scheduler is a modular plugin with different possible allocation algorithms that find a suitable mapping from processors in the dataflow to resources, based on the capacity and QoS. 

The manager then contacts a \emph{deployer} module that enacts the mapping of processors to resources, connecting them across different resources, and starting the dataflow execution. For this, it invokes a Platform Service running on each resource that in turn interfaces with the local dataflow engine for processor deployment. Once successfully started, the manager assigns a UUID to the dataflow, registers the dataflow JSON and its resource mapping with the registry, and returns the UUID to the user. 
This UUID can be used to later manage the dataflow, say, to stop it. In this case, the user again contacts the master which spawns a manager that then retrieves the dataflow's state from the Resource Directory. It then works with the deployer to contact the platform services on the resources in which this dataflow's processors are running, stops and undeploys them, and updates the dataflow's state in the registry.

\noindent \textbf{Platform Service and Distributed Orchestration.} 
The container or VM that will host the application runs a \emph{platform service} for managing the dataflow orchestration on it. Depending on the resource availability and sharing allowed between dataflows of the same or different tenants, each container can run all or parts of one or more dataflows. We use \emph{Apache NiFi}, a light-weight engine designed for interactively composing modular processors and executing a dataflow on a single machine, as our base dataflow orchestration engine. 
NiFi's native data model is a \emph{FlowFile}, which is an in-memory reference to a collection of bytes, which may be persisted to disk as one or more files, along with attributes describing it. We treat a FlowFile as a micro-batch, and provide \emph{data model wrappers} to/from streams and files from FlowFiles. 

Processors are user-defined Java logic that can access the attributes of a FlowFile, and its contents as a byte stream, and likewise generate new FlowFiles that are passed to downstream processors in the dataflow by the engine. NiFi offers limited support for distributed devices. Instances on different machines can pass FlowFiles between their processors by manually defining and wiring a \emph{remote process group (RPG)}. RPGs can use HTTP or a binary protocol to push FlowFiles downstream or pull FlowFiles from upstream processors.

We extend NiFi in several ways to meet the listed desiderata. Our platform service uses the NiFi APIs to programmatically deploy and execute fragments of one or more dataflows in a single engine. Since the resource scheduler may map different processors in the dataflow to different resources, each NiFi engine may have only a subset of it. E.g., in Fig~\ref{fig:arch}, $N_1,T_1$ and $S_1$ are part of the same dataflow but placed in a Pi, a TX1 and a VM. We treat NiFi as a local orchestration container for multiple fragments. The deployer coordinates among different NiFi instances by automatically introducing RPGs at the edge-cut of the dataflow graph that span resources. While RPGs currently push FlowFiles downstream, knowledge of network restrictions 
can be used to decide if an upstream RPG is a client (push) or a server (pull) to the downstream RPG. This ambi-directionality allows the platform to execute dataflows between resources even when they one is behind a firewall. 

We further introduce specialized \emph{runtime wrapper processors}, as discussed in \S~\ref{sec:model}, for native support for external runtime engines. Specifically, we support Edgent for in-memory CEP, TensorFlow for deep learning models using CPU and GPGPU, and Spark and Storm for stream and batch processing of Big Data. While the Edgent processor operates within NiFi, TensorFlow is forked as a process on the local device from the processor. Both these also use data model wrappers, as shown in Fig.~\ref{fig:model}. The Storm and Spark processors also require support within the native dataflow. Specifically, we have  source and sink tasks of the Storm or Spark dataflow interface with the RPGs of NiFi to transfer the FlowFiles between the different engines, with an optional data model wrapper. Users just provide the external engine's dataflow logic to our runtime wrapper processors, which then launches and interacts with it transparently.

Lastly, we provide first-class support for dynamic migration of the dataflows at execution time to adapt to external conditions. Dataflow \emph{rebalancing} refers to the process of migrating running processors from the resources they are present in to different ones. While rebalance is explicitly triggered by the user now, it is possible to have the app manager periodically check the QoS of the application and pro-actively initiate this rebalancing. 
A user's call to the master to rebalance spawns a manager thread to query the current dataflow and mapping from the registry, and pass it to the scheduler to get an updated resource allocation. The manager then contacts the deployer with the old and the new mappings, which performs a graph ``diff'' to identify processors that need to be migrated. It then pauses the processors that are being migrated and their adjacent ones, migrates the relevant processors, introduces/removes RPGs at the new/old boundaries, and rewires the processors before resuming them. During this time, unpaused processors continue to execute, though inputs to paused processors will queue.

\section{Evaluation and Results}
\label{sec:results}We evaluate the \echo architecture and implementation for real-world IoT data\-flows that support the Smart Community use-case we motivated earlier. We deploy \echo on an \emph{IoT testbed} at our Indian Institute of Science (IISc) campus in Bangalore with the following setup of local Edge and Fog devices within 2 network hops on the private network, complemented by Microsoft Azure VMs at 2 data centers. The Platform Master and Resource Directory services run on an exclusive DS1 VM each, while the rest are available for deploying applications. These are described in Table~\ref{tbl:resources}.

\begin{table}[t]
\centering
\footnotesize
\caption{Set of resources used in the IoT testbed for evaluation}\label{tbl:resources}
 	\begin{tabular}{c|c|p{3.4cm}|c|c|c}\hline
 		\bf Resource & \bf Count & \bf CPU/GPU & \bf RAM & \bf NIC & \bf Location \\ \hline
 		 \hline
Pi 3B Edge & 10 & 900MHz ARM A53 64bit, 4~cores & 1GB & 100Mbps & IISc \\ \hline
Pi 2B Edge & 2 & 900MHz ARM A7 32bit, 4~cores & 1GB & 100Mbps & IISc\\ \hline
TX1 Fog & 1 & 1.75GHz ARM A57 64bit, 4~cores; Nvidia Maxwell, 256 CUDA cores & 4GB & 1Gbps & IISc \\ \hline
Softiron Fog & 1 & 2GHz AMD A1100 (ARM A57) 64bit, 8~cores & 16GB & 1Gbps & IISc \\ \hline
DS1 v2 VM & 4 & 2.4GHz Intel Xeon E5 v3, 1~core & 3.5GB & 2$\times$1Gbps & South India\\\hline
NC6 VM & 1 & 2.6GHz Intel Xeon E5 v3, 6~cores; Nvidia K80, 4992~CUDA cores & 56GB & 1Gbps & US East\\
 		 \hline
 	\end{tabular}
 \end{table}


	
\begin{figure}[t]
	\centering
	\subfloat[ETL]{
		\includegraphics[width=0.48\textwidth]{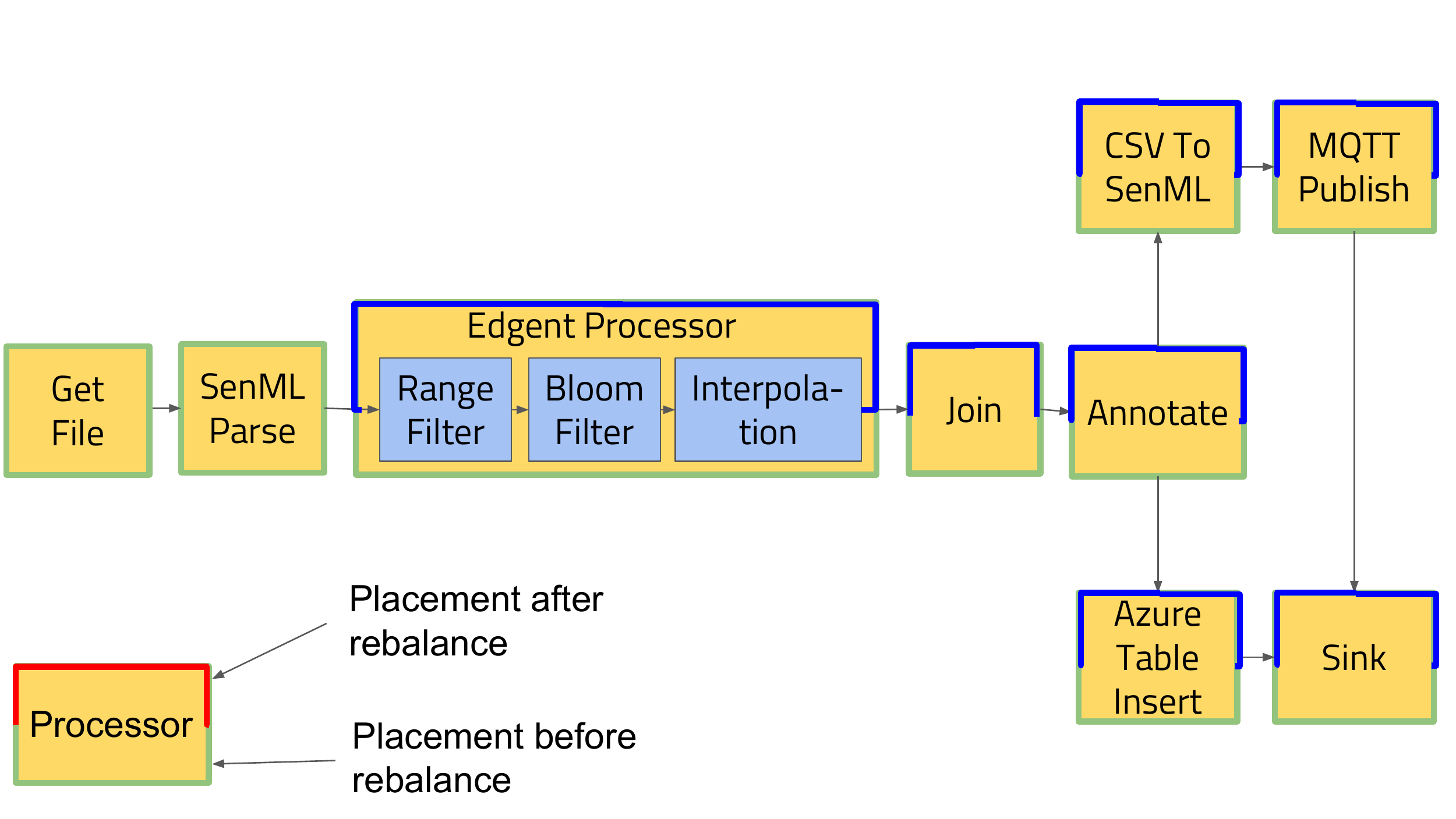}
		\label{fig:df:etl}
	}
	\subfloat[YOLO]{
		\includegraphics[width=0.48\textwidth]{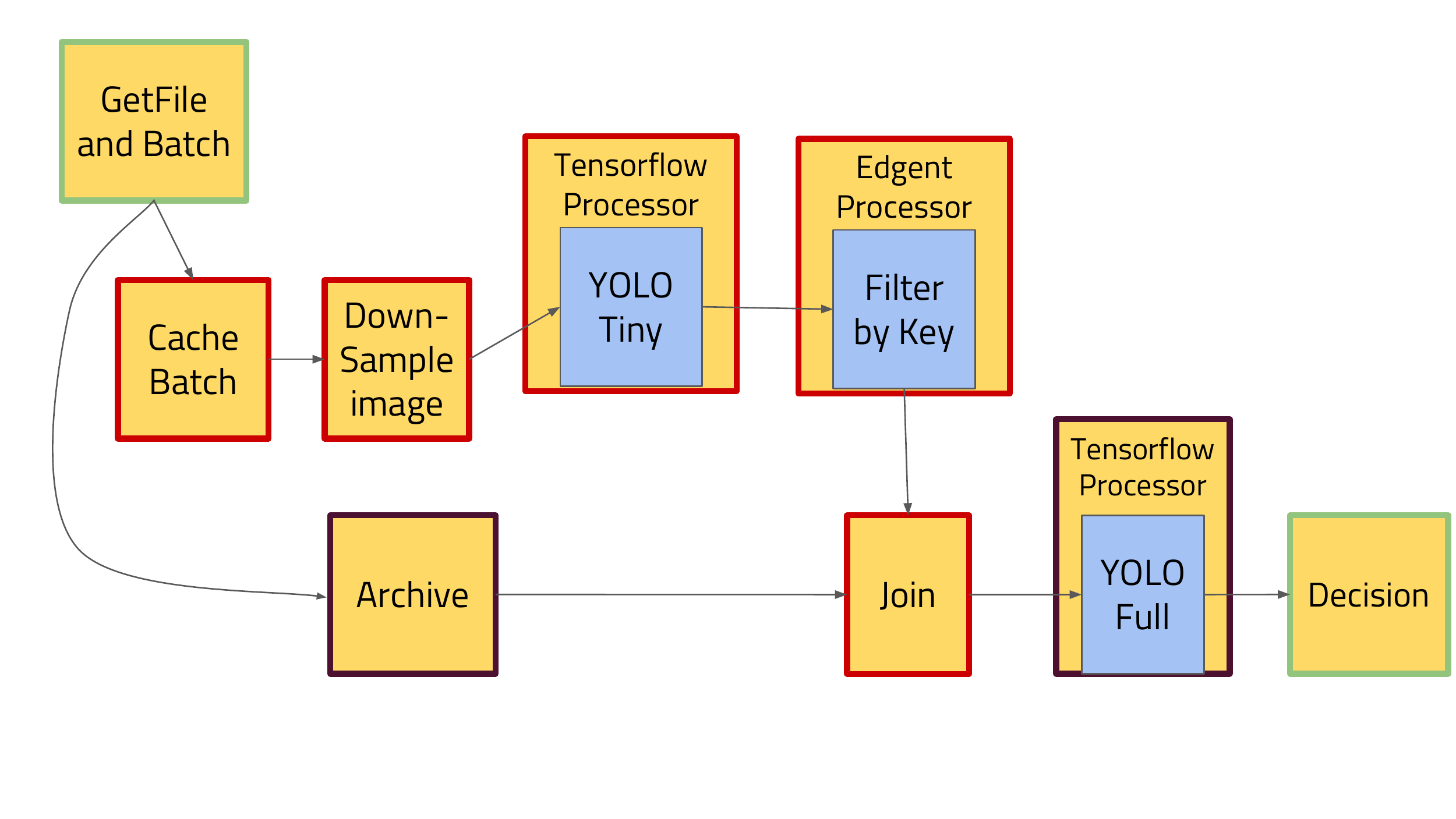}
		\label{fig:df:yolo}
	}\\
	\subfloat[STATS]{
		\includegraphics[width=0.48\textwidth]{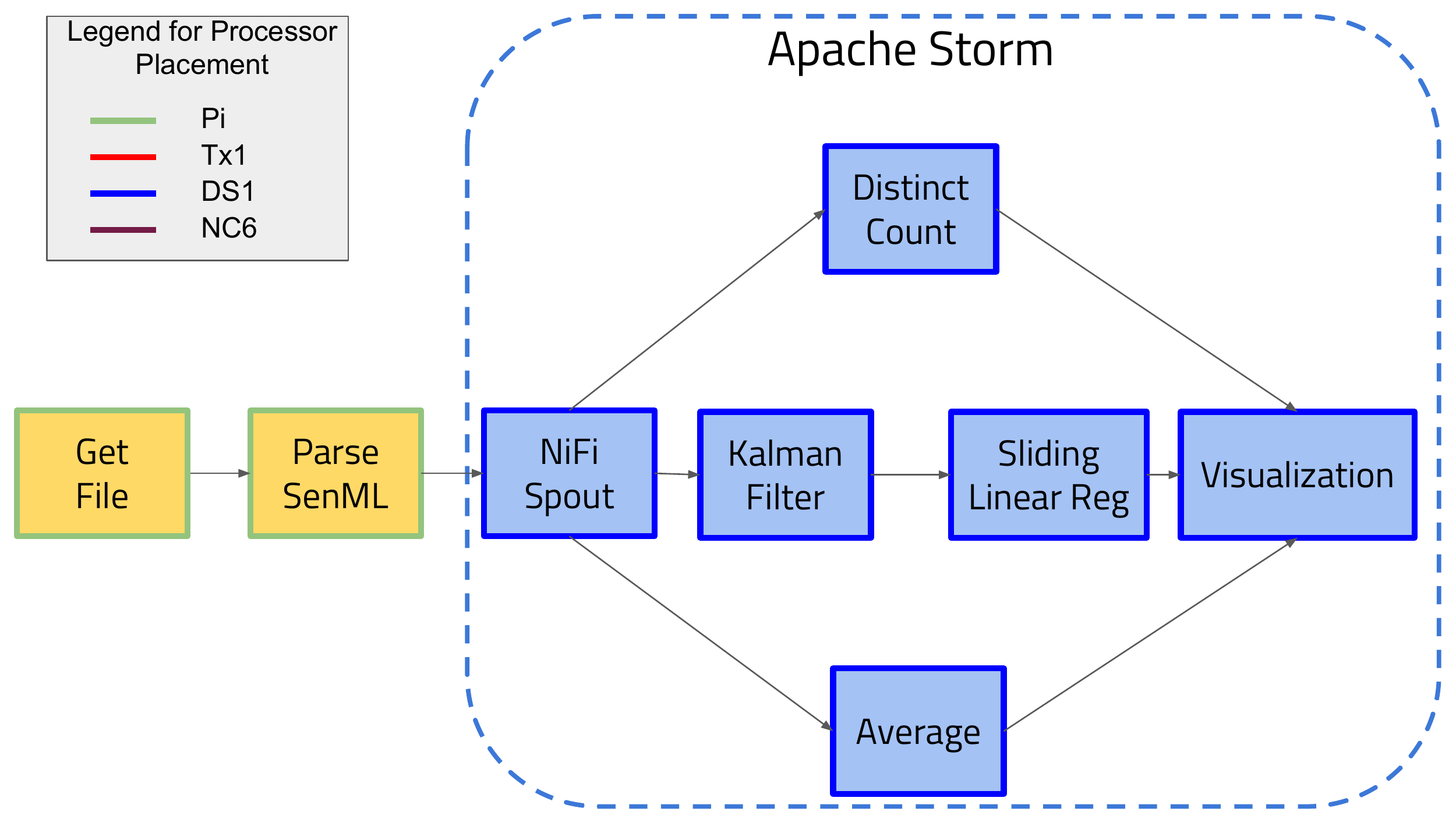}
		\label{fig:df:stats}
	}\hfill
\subfloat[NiFi UI for Live Visualization]{
		\includegraphics[width=0.37\textwidth]{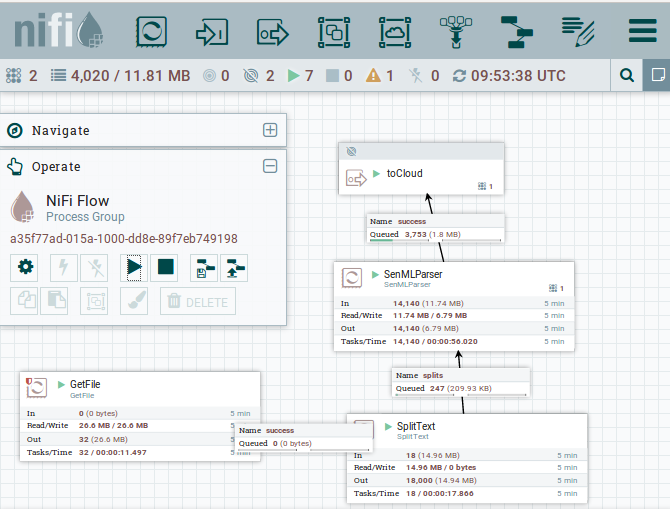}
		\label{fig:df:stats}
	}
	\caption{Smart City dataflows used in evaluation}
	\label{fig:dataflows}
\end{figure}

The three IoT application dataflows used in the validation are shown in Fig.~\ref{fig:dataflows} and summarized in the Table~\ref{tbl:dataflows}. These are based on real-world data processing and analytics for smart utility and traffic surveillance scenarios. 
\begin{table}[t]
\centering
\footnotesize
\caption{Set of dataflows used in the evaluation}\label{tbl:dataflows}
\footnotesize
 	\begin{tabular}{c||p{1.5cm} p{2cm} p{2cm} p{2cm}}\hline
 		\bf Dataflow & \bf Input & \bf Platforms & \bf Data Model & \bf Resources \\ \hline \hline
 		ETL & NYC Taxi & NiFi, Edgent & $\mu$-batch, Stream & Pi, VM$_{\text{DS1}}$, S'iron \\ \hline
 		YOLO & Pedestrian Video~\cite{eth_biwi_00534} & NiFi, T'Flow, Edgent  & $\mu$-batch, Stream, File & Pi, TX1, VM$_{\text{NC6}}$ \\ \hline
 		STATS & NYC Taxi & NiFi, Storm & $\mu$-batch, Stream & Pi, VM$_{\text{DS1}}$ \\
 		 \hline
 	\end{tabular}
 \end{table}
 
The Extract Transform Load (\emph{ETL}) dataflow performs data pre-processing and cleaning of sensor observation streams, such as smart grids and environmental sensing, before archiving then to Cloud storage~\cite{DBLP:journals/corr/ShuklaCS17}. It parses the input SenML micro-batch in NiFi, streams each observation to Edgent for filtering, outliers detection, and interpolation using its built-in CEP tasks, annotates it as micro-batches back in a NiFi processor before publishing to an MQTT pub-sub broker and to an Azure NoSQL table concurrently 
We run it on NY Taxi event streams~\cite{DBLP:journals/corr/ShuklaCS17}. The tasks initially run on 4 Pi devices, but are rebalanced and migrated mid-way to also use 2 Cloud VMs.

\emph{YOLO}~\cite{DBLP:journals/corr/RedmonF16} is a deep convolutional neural network (CNN) for TensorFlow to classify pedestrians in frames of traffic videos. We use it for both pre and post processing, on edge with low latency and on Cloud with high accuracy. In our dataflow, video segments are in parallel archived on a Pi, and also downsampled to $416\times416$~px for efficient detection using a YOLO Tiny model on the TX1. YOLO returns a text label and bounding box, which are streamed as tuples to an Edgent processor to detect patterns of interest, say more than 5 people in a frame. Upon a match, we push the corresponding video frames at original resolution ($2.1\times$ larger) to a Cloud GPU VM for accurate classification by a YOLO Full TensorFlow model. A match triggers an alert for further action.

Lastly, a statistical analytics dataflow (\emph{STATS}) is an IoT application~\cite{DBLP:journals/corr/ShuklaCS17} that performs streaming analysis over events with high velocity. It concurrently does a Kalman filter smoothing and linear regression, windowed aggregation, and distinct count of sensors, which are then plotted and the images zipped for publishing online. These tasks are designed as a Storm topology that run on Cloud VMs, with a NiFi processor passing it event batches from the edge, and receiving the response. 
As we can see, these three dataflows capture real scenarios that cannot be adequately met by a single dataflow platform, a single data model or a single type of device, highlighting the value of \echo.

	\begin{figure}[t]
		\centering
\subfloat[ETL Throughput and CPU\% with Rebalance]{
		\includegraphics[width=.32\textwidth]{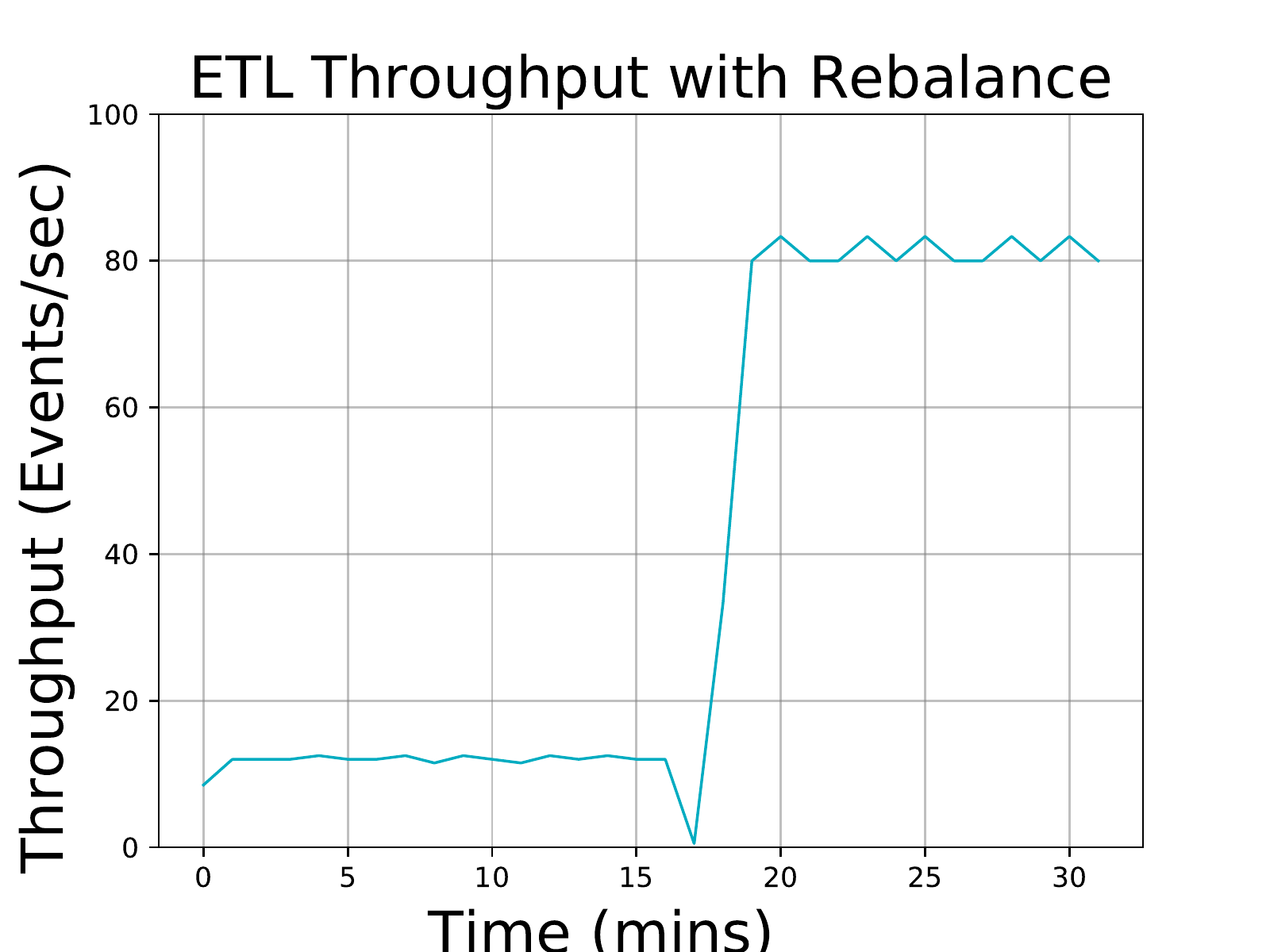}
		\includegraphics[width=.32\textwidth]{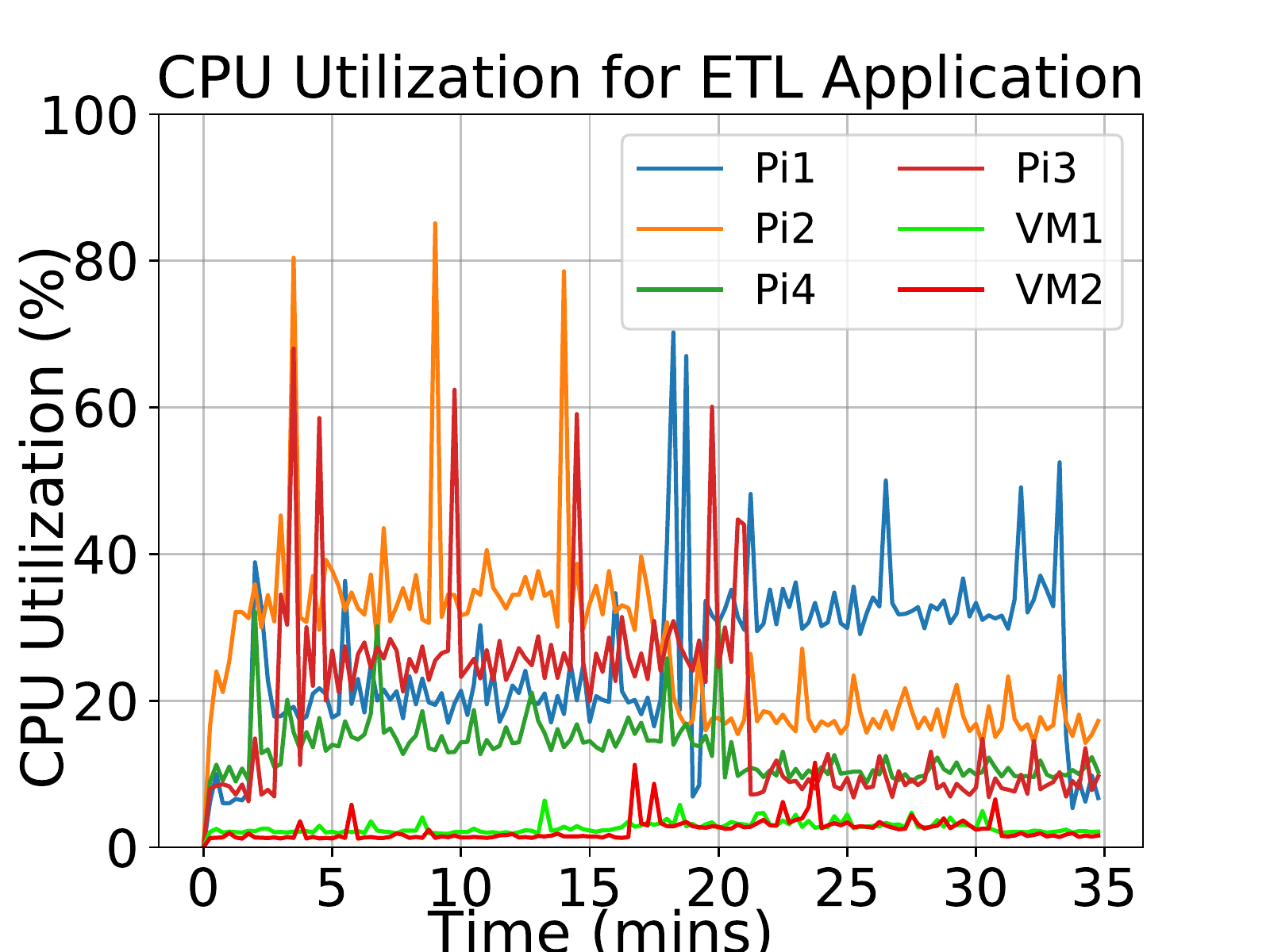}
		\label{fig:res:etl}
}
\subfloat[STATS Throughput]{
	\includegraphics[width = 0.31\textwidth]{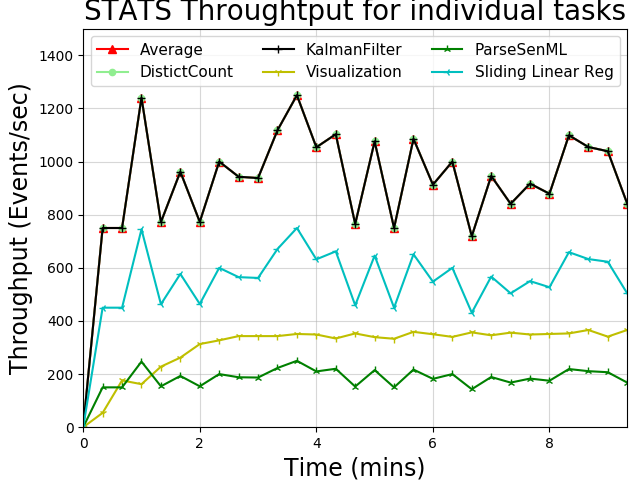}
		\label{fig:res:stats}
}\\
	\subfloat[YOLO CPU\%, Memory\% and Frame Rate supported by Tiny and Full models]{
	\includegraphics[width=0.32\textwidth]{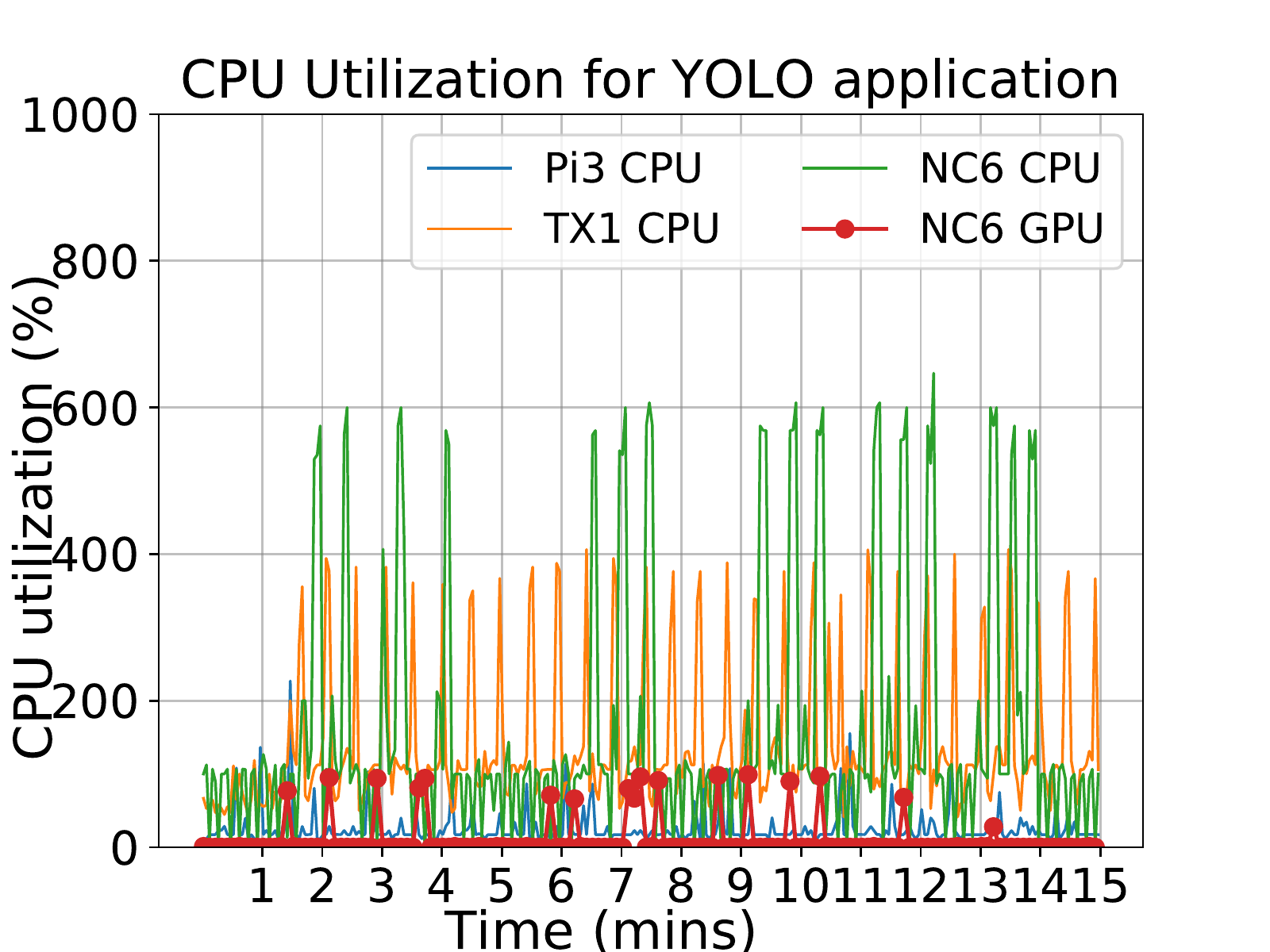}%
	\includegraphics[width=0.32\textwidth]{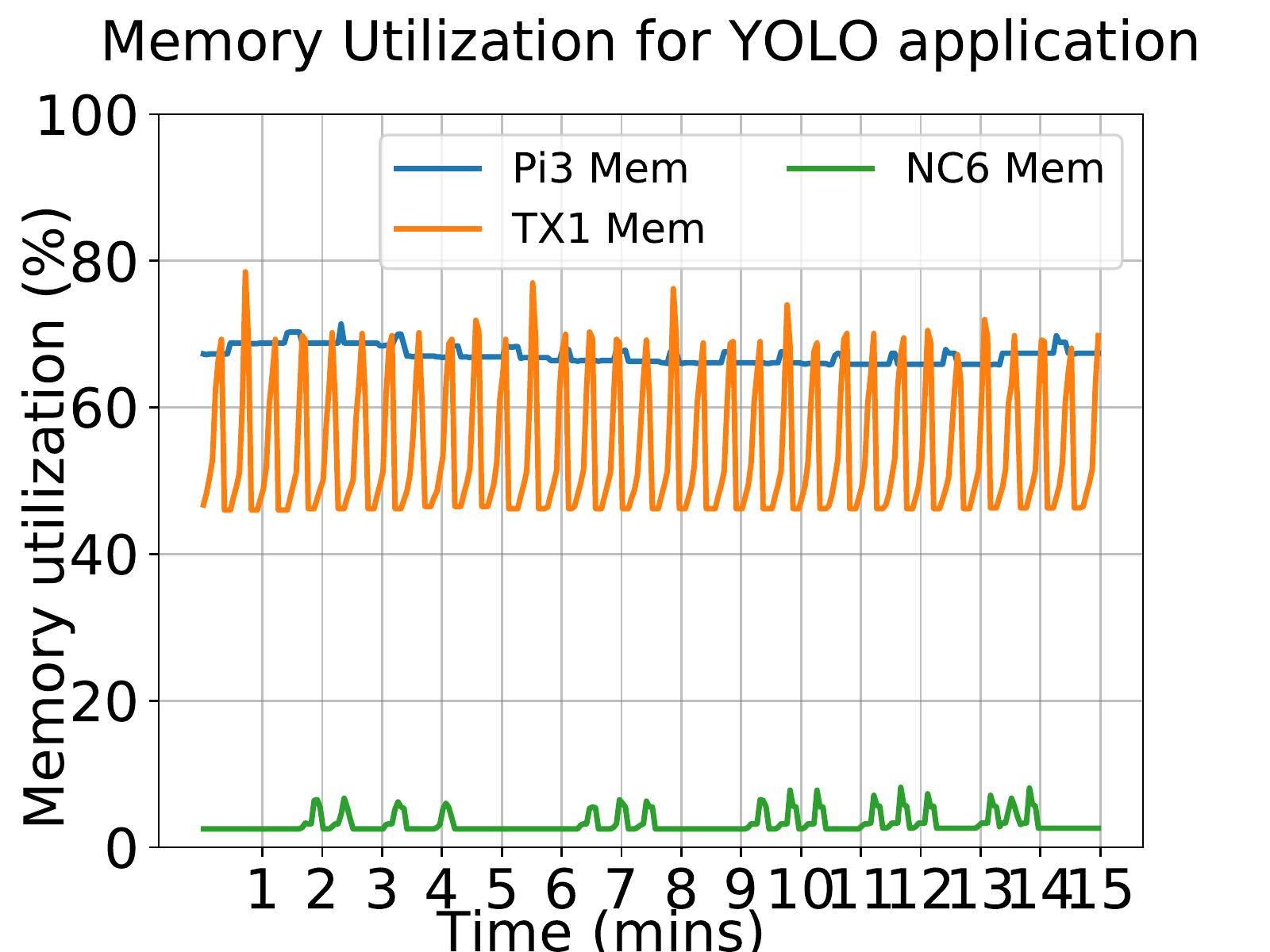}%
	\includegraphics[width=0.32\textwidth]{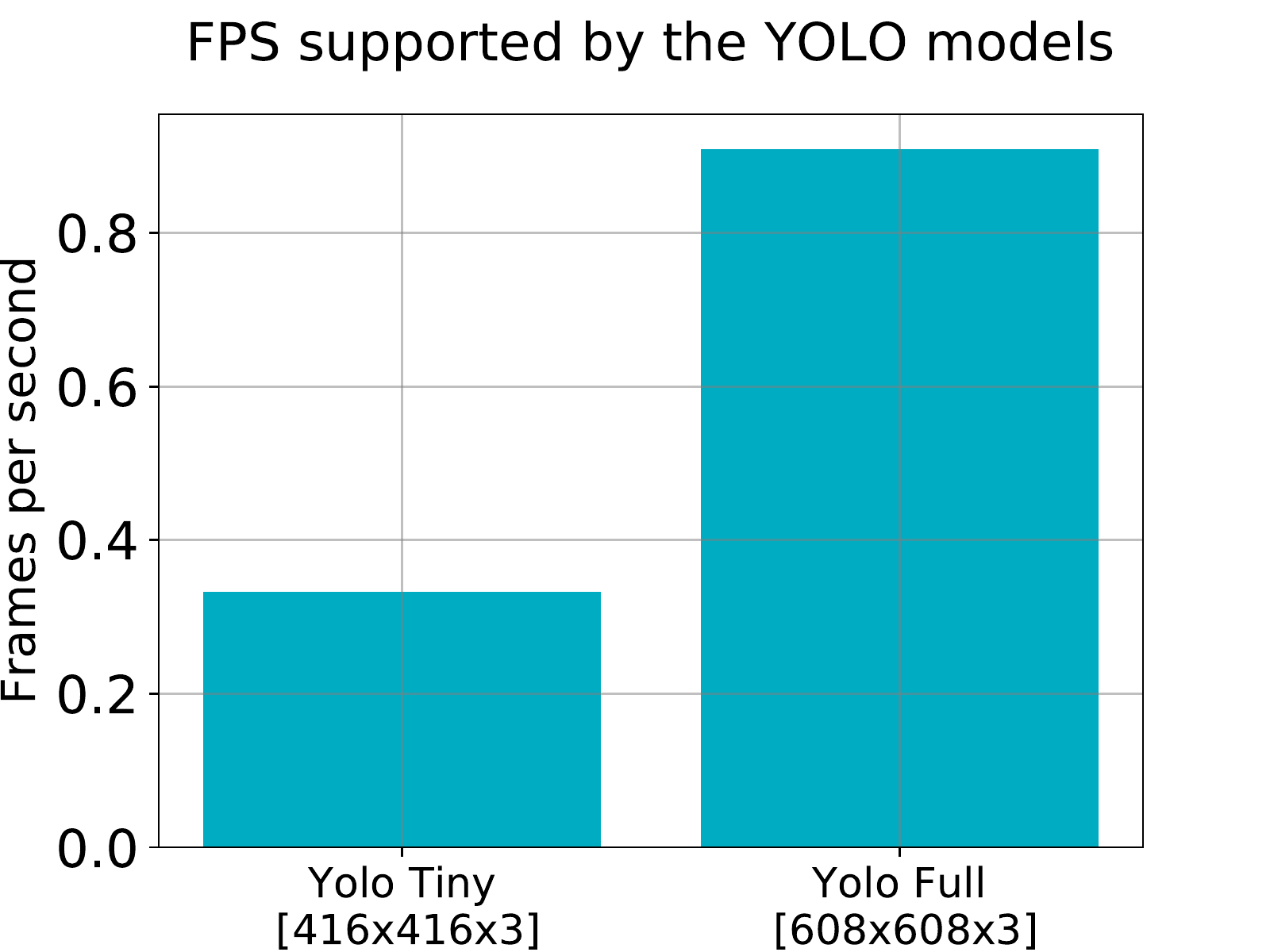}%
	\label{fig:res:yolo}
}
\caption{Results for ETL, YOLO and STATS dataflows}
\end{figure}

\noindent\textbf{Results.} We deploy the dataflows on the IoT testbed devices and the VMs using a custom scheduler, and offer representative samples of the performance results upon running them continuously. Fig.~\ref{fig:res:etl} shows the output event rate, and CPU\% on each active device for ETL across time. In the first half, we schedule the processors only on Pi's but initiate a dynamic rebalance at the mid-point to additionally use 2 VMs. As we see, the supported event rate jumps from 15~events/sec to 80~events/sec, with a brief dip while the migration occurs. We see a corresponding change in the CPU\% as well, with the usage on Pi1 increasing as it is retained after rebalance while other Pi's dropping low, and the VM usage marginally increasing. Despite having more cores, the Pi's have 3x slower clockspeeds, and hence offer limited throughput.

The batch behavior of YOLO clearly shows in its CPU\% and Memory\% plots over time in Fig.~\ref{fig:res:yolo}, with the spikes coinciding with a micro-batch or file being processed by NiFi or TensorFlow. This happens across CPU, GPU, Pi, TX1 and VM, but is more prominent on TX1 since it is the most stressed resource when running the YOLO Tiny model. The frame-rate supported by YOLO Tiny on TX1 is $\frac{1}{3}^{rd}$ that of Yolo Full on NC6, despite having $\frac{1}{2}$ the image size. The NC6 VM has a much faster GPU and spare capacity, indicating that a single GPU VM can service multiple video streams to complement the Fog servers.

We report the throughput at each NiFi or Storm task in the STATS dataflow in Fig.~\ref{fig:res:stats}. We can see that the use of Storm helps support high input rates of over 1000~events/sec. The variation in rates is due to the selectivity of different tasks, that can produce more or fewer events than what they consume. The rates are also smoother than YOLO, reflecting the streaming data model used.

\section{Related Work}
\label{sec:related}
The lack of middleware for IoT and edge-computing is well recognized~\cite{icsoc:2016:middleware,varshney:icfec:2017,epema:edge,simmhan:iot:2017}, even as the growing deployment of such devices and applications use bespoke solutions. 
\cite{DBLP:journals/corr/MineraudMST15a} offers a gap analysis of IoT platforms, several of which \echo addresses including the use of Edge, Fog and Cloud resources, easing development of distributed dataflow applications, and automating the environment setup. 

Many proprietary and open source projects have recently evolved. 
\emph{Eclipse Kura}~\cite{kura} is a Java-based gateway management project for Linux edge devices that allows application deployment using OSGi containers. But it does not support dataflow composability within or across devices. \emph{VMWare's Liota}~\cite{liota} is a similar Python-based management stack with sensor, pub-sub and Cloud service bindings that can run local applications on a device. A proprietary version also integrates with their data center infrastructure fabric management suite. Both of these complement \echo's PaaS layer and can form the IaaS layer.

Cloud providers like Amazon AWS and Microsoft Azure have extended some of their Cloud features to tightly integrate with edge devices as well. \emph{Amazon's GreenGrass}~\cite{greengrass} is an IoT SDK that allows users to deploy AWS Lambda functions on edge devices, and use MQTT for coordination. They also offer bindings with AWS Cloud services like S3 and DynamoDB. \emph{Azure IoT Edge}~\cite{azureiot} has a similar goal. In both cases, the SDK offer some programming and management capabilities on the edge but push analytics to their Cloud services. Composability, support for external Edge runtimes, hybrid data model, etc. are non-goals.


\emph{Apache Edgent} as we saw offers a CEP platform for Edge devices. This is designed as a stand-alone embedded library rather than for composable dataflows. \emph{Node.RED} is similar to NiFi in providing interactive dataflow composition across devices using a Node.js server. But its features are restricted, supporting only JavaScript tasks, although it is more light-weight. \emph{MiNiFi} is a light flavor of NiFi that supports C++ and embedded platforms, but trims many of NiFi's features like online deployment and dynamic migration ability. We attempt to balance features and footprint in \echo, and can leverage these alternative dataflow engines in future to complement NiFi.

IoT Middleware is an active research area as well. 
The \emph{MiMove project}~\cite{icsoc:2016:middleware} has proposed an SOA architecture for mobile IoT, with a focus on the functional scalability. A novel probablistic registry allows low-latency approximate queries for registered sensing and actuation services. It does static scheduling of streaming service dataflows using the Dioptase middleware~\cite{billet:2014:mobihoc}, and interfacing across heterogeneous IoT protocols using an Enterprise Service Bus (ESB). \echo in contrast supports hybrid data models -- a higher level abstraction than protocols, offers richer composition semantics including delegating to external engines, and uses point-to-point communication between tasks (push and pull) rather than a central ESB. Our Hypercat registry is simpler but based on BSI standards, and can be replaced or federated for scaling. Advanced scheduling algorithms~\cite{ghosh:2016} or device mobility is not a focus in this paper, but future work.

\cite{hong:mcc:2013} has proposed a programming model for composing IoT applications across mobile, Fog and Cloud layers. They consider a multi-way 3-level dataflow model with computation starting in the Cloud, elastic resources acquired in the Cloud and Fog, and communication possible between all 3 layers. Each edge has one Fog parent based on spatial proximity, that may be reassigned. 
While a useful abstraction, their strictly hierarchical resource and dataflow model are much more restrictive that our use of any network topology and a directed graph as dataflow. 
Theirs effectively degenerates to a client-server model. 

\cite{DeConinck:2015:DDA:2836127.2836130} is a middleware framework that is built specifically for feed-forward execution of distributed neural networks across multiple IoT devices, with variable latency to the Cloud and privacy being the rationale for the distribution. The layers of the neural networks are represented as modules which are composed in the form of a dataflow. It is also optimized for a single sample feed forward execution unlike most other frameworks which are optimized for batch processing. During deployment, however, it fails to take into consideration statics such as CPU/Memory Utilization nor has provisions for obtaining the same. \cite{Ichinose:2017:PPD:3022227.3022323} is a pipeline-based distributed processing extension for Caffe deep learing framework. Two custom layers \textit{Source} and \textit{Sink} were added to the Caffe framework that allow splitting a single neural network across Edge and Cloud. The extension allows for both feed-forward as well as back-propogation, thus enabling learning as well. The work also explores varing the parameters and it's effect on the required bandwidth and learing accuracy. However, the point of splitting the neural network is statically determined and even the placement is manual without taking into consideration any heuristics.

\emph{Mobile Clouds} are precursors to IoT where mobile phones off-load applications to Cloud resources.
In~\cite{yang2013framework}, mobile data stream applications are dynamically partitioned for computation across mobile devices and Cloud. They propose a genetic algorithm for the partitioning to maximize throughput and adapt to changing devices load. 
They are limited to mobile data stream applications rather than dataflow or hybrid data models 
We also support Fog resources, native runtime engines and dynamic migration of tasks among the resources. 
The Hybrid Mobile Edge Computing (HMEC) architecture~\cite{reiter2017hybrid} uses edge devices for mobile applications. They use a peer-to-peer (P2P) approach of both proximate and distant edge devices, and perform  \emph{method-based offloading} to improve performance and reduce energy usage. 
Similarly,~\cite{chun2011clonecloud} offloads tasks to the Cloud using RPC with static analysis and dynamic profiling of mobile applications.  
It maintains a complete device clone in the Cloud, which can be costly. These are designed for monolithic existing mobile applications rather than \emph{ad hoc} dataflow composition, and neither consider a service paradigm or Fog servers. 

In~\cite{saurez2016incremental}, a C++ programming framework is proposed for Fog that provides APIs for resource discovery, migration, communication and QoS-aware incremental deployment of fog cells and services via containerisation. However they do not consider diverse applications such as streams, microbatches and dataflows altogether.  They have used vehicular traffic simulation application for experimental evaluation by setting up a real-world fog landscape using docker containers deployed on several servers.
	


\emph{P2P frameworks} like Seti@home\cite{seti} have targeted the use of idle compute capacity in desktops. 
However, some of the inherent P2P characteristics are missing in an IoT scenario.  
Device churn is a major factor in P2P but less so for infrastructure IoT, or even mobile devices that are typically within cell communication.  
This, coupled with the growth of global Cloud data centers, make it feasible for centralized services for coordination. Dataflow composition is also a non-goal for such P2P systems that typically use a task-queue model for opportunistic computing. 




\section{Conclusions}
\label{sec:conclusions}
In this paper, we motivate the gaps and propose the requirements for a middleware platform to compose and orchestrate dataflows across Edge, Fog and Cloud resources to support IoT applications. We present \echo, a service-oriented platform that addresses these design requirements. It includes novel features such as dataflow composition using hybrid data models like streams, micro-batch and files; inherent support for external runtime engines like Edgent, Tensorflow, Storm and Spark; and dynamic migration of dataflow tasks across distributed resources. \echo also offers basic capabilities of dataflow orchestration using NiFi, a standards-compliant registry, and containerization for light-weight resource sharing. We empirically validate and present results for three real-world IoT applications that exercise the various features of \echo.

\section{Future Work}

This work addresses the highlighted gaps, but much more remains in this emerging area. It is worth examining a more decentralized decision making for deployment and scheduling rather than execute a single centralized cloud Master. This will be essential for scaling to millions of devices, and the Fog can play a role here. This goes hand in hand with work on creating optimal scheduling algorithms that can are practical to run on the platform. We should also explore scalable federated catalogs as data sources and replicas are included in the registry, and have devices actively join, leave, and update their state. Auditing, billing and tracking of data will be come important. NiFi inherently supports provenance collection and this can be leveraged to replay historic streams and data. Migration of processors that hold state is another useful feature that would be taken up. The ability to detect faulty network connections and compute resources, and take action accordingly, would be required to make the platform more stable.


\section*{Acknowledgments}
 The authors would like to thank Microsoft Azure and NVIDIA to access to Cloud and TX1 resources. We would also like to thank Venkatesh Babu and Avishek from the VAL lab at IISc for inputs on the video analytics deep-learning model.


\bibliographystyle{plain}


\end{document}